\documentclass[showpacs,aps,twocolumn]{revtex4}
\usepackage{graphicx}
\begin{document}
%%%%%%%%%%%%%%%%%%%%%%%%%%%%%%%%%%%%%%%%%%%%%%%%%%%%%%%%%%%%%%%%%%%%%%%%%%%%%%%%
%%%%%%%%%%%%%%%%%%%%%%%%%%%%%%%%%%%%%%%%%%%%%%%%%%%%%%%%%%%%%%%%%%%%%%%%%%%%%%%%
\title{Stationary Light Pulses without Bragg Gratings}
\author{Yen-Wei Lin, Hung-Chih Chou, Thorsten Peters, Wen-Te Liao, Hung-Wen Cho, Pei-Chen Guan, and Ite A. Yu}
\email{yu@phys.nthu.edu.tw}
\affiliation{Department of Physics, National Tsing Hua University, Hsinchu 300, Taiwan, Republic of China}
\date{August 20, 2008}
%%%%%%%%%%%%%%%%%%%%%%%%%%%%%%%%%%%%%%%%%%%%%%%%%%%%%%%%%%%%%%%%%%%%%%%%%%%%%%%%
%%%%%%%%%%%%%%%%%%%%%%%%%%%%%%%%%%%%%%%%%%%%%%%%%%%%%%%%%%%%%%%%%%%%%%%%%%%%%%%%
\begin{abstract}
The underlying mechanism of the stationary light pulse (SLP) was identified as a band gap being created by a Bragg grating formed by two counter-propagating coupling fields of similar wavelength. Here we present a more general view of the formation of SLPs, namely several balanced four-wave mixing processes sharing the same ground-state coherence. Utilizing this new concept we report the first experimental observation of a bichromatic SLP at wavelengths for which no Bragg grating can be established. We also demonstrate the production of a SLP directly from a propagating light pulse without prior storage. Being easily controlled externally makes SLPs a very versatile tool for low-light-level nonlinear optics and quantum information manipulation.
\end{abstract}
%%%%%%%%%%%%%%%%%%%%%%%%%%%%%%%%%%%%%%%%%%%%%%%%%%%%%%%%%%%%%%%%%%%%%%%%%%%%%%%%
%%%%%%%%%%%%%%%%%%%%%%%%%%%%%%%%%%%%%%%%%%%%%%%%%%%%%%%%%%%%%%%%%%%%%%%%%%%%%%%%
\pacs{42.50.Gy, 32.80.Qk}
%Effects of atomic coherence on propagation, absorption, and amplification of light.
%Coherent control of atomic interactions with photons.
\maketitle
%%%%%%%%%%%%%%%%%%%%%%%%%%%%%%%%%%%%%%%%%%%%%%%%%%%%%%%%%%%%%%%%%%%%%%%%%%%%%%%%
%%%%%%%%%%%%%%%%%%%%%%%%%%%%%%%%%%%%%%%%%%%%%%%%%%%%%%%%%%%%%%%%%%%%%%%%%%%%%%%%
Based on the effect of electromagnetically induced transparency (EIT) \cite{EitRv1, EitRv2}, a light pulse can be stored in and subsequently released from a medium \cite{StopLtTh, HauStopLt, WalsworthStopLt}. During the storage, the electromagnetic component of the light is completely converted into a spin excitation, i.e., ground-state coherence of the medium. Thus, there is no light in the medium during the storage which, however, is required for nonlinear optical interactions. To actually stop a light pulse while maintaining an electromagnetic component, i.e., to create a stationary light pulse (SLP) has been proposed by Andr\'{e} and Lukin \cite{SLP3} and experimentally demonstrated by Bajcsy {\em et al.} \cite{SLP1}. A probe pulse propagating inside a medium prepared by EIT is first converted into a spin excitation of the medium with the light storage technique. Then, a standing wave formed by two counter-propagating coupling fields of similar wavelength is applied during the retrieval which leads to the creation of a SLP. This is explained by the periodic modulation of the absorption produced by the standing wave that acts like a Bragg grating, or in other words, causes the medium to act like a photonic band gap medium and prohibits propagation of the probe pulse.

Here now, we would like to present a more general point of view towards the underlying mechanism of SLPs and start the discussion with the relevant optical Bloch equations of the density matrix operator $\rho$ of the system and the Maxwell-Schr\"odinger equation of the probe field:
\begin{eqnarray}
\label{one1}
    \frac{\partial \rho_{21}}{\partial t}  = 
        \frac{i}{2} \Omega^{*}_{c}\rho_{31}-\gamma\rho_{21}, \\
\label{one2}
    \frac{\partial \rho_{31}}{\partial t}  = 
        \frac{i}{2} \Omega_{p} +\frac{i}{2} \Omega_{c}\rho_{21}
        -\frac{\Gamma}{2}\rho_{31}, \\
\label{one3}
    \frac{1}{c}\frac{\partial \Omega_p}{\partial t}
        +\frac{\partial \Omega_p}{\partial z}
         =  i \frac{\alpha \Gamma}{2L} \rho_{31},
\end{eqnarray}
where $\Omega_p$ and $\Omega_c$ are the Rabi frequencies of the probe pulse and the coupling field, $\gamma$ is the relaxation rate of the ground-state coherence, $\Gamma$ is the spontaneous decay rate of the excited state, and $\alpha$ and $L$ are the optical density and the length of the medium, respectively. Next, we modify the equations above to allow us to perform calculations on SLPs. As there are forward- and backward-propagating probe pulses and coupling fields when SLPs are considered, we replace $\Omega_c$, $\Omega_p$, and $\rho_{31}$ by $\Omega^+_c {\rm e}^{ik^+_c z} +\Omega^-_c {\rm e}^{-ik^-_c z}$, $\Omega^+_p {\rm e}^{ik^+_p z} +\Omega^-_p {\rm e}^{-ik^-_p z}$, and $\rho^+_{31} {\rm e}^{ik^+_p z}+\rho^-_{31} {\rm e}^{-ik^-_p z}$, respectively. Here, $k_p^{\pm}$ and $k_c^{\pm}$ are the wave vectors of the probe and coupling fields in the $\pm z$-direction, respectively. By neglecting the fast oscillating terms containing ${\rm e}^{i(k^+_p+k^-_c) z}$ and ${\rm e}^{-i(k^-_p+k^+_c) z}$ as well as considering $k^+_p \approx k^+_c$ and $k^-_p \approx k^-_c$, Eqs.~(\ref{one1})-(\ref{one3}) become
\begin{eqnarray}
\label{two1}
  \frac{\partial \rho_{21}}{\partial t}  = 
        \frac{i}{2} (\Omega^{+}_{c})^* \rho^+_{31}
        +\frac{i}{2} (\Omega^{-}_{c})^* \rho^-_{31}
       -\gamma\rho_{21}, \\
\label{two2}
    \frac{\partial \rho^+_{31}}{\partial t}  = 
        \frac{i}{2} \Omega^+_{p} +\frac{i}{2} \Omega^+_{c}\rho_{21}
        -\frac{\Gamma}{2}\rho^+_{31}, \\
\label{two3}
    \frac{\partial \rho^-_{31}}{\partial t}  = 
        \frac{i}{2} \Omega^-_{p} +\frac{i}{2} \Omega^-_{c}\rho_{21}
        -\frac{\Gamma}{2}\rho^-_{31}, \\
\label{two4}
    \frac{1}{c}\frac{\partial \Omega^+_p}{\partial t}
        +\frac{\partial \Omega^+_p}{\partial z}
         =  i \frac{\alpha \Gamma}{2L} \rho^+_{31}, \\
\label{two5}
    \frac{1}{c}\frac{\partial \Omega^-_p}{\partial t}
        -\frac{\partial \Omega^-_p}{\partial z}
         =  i \frac{\alpha \Gamma}{2L}\rho^-_{31}.
\end{eqnarray}
Solving Eqs.~(\ref{two1})-(\ref{two5}) numerically for a timing of $(\Omega^+_c)^2$ and $(\Omega^-_c)^2$ as shown in Fig.~1(a), the probe field is obtained as a function of position and time as represented by $(\Omega^+_p)^2$ and $(\Omega^-_p)^2$ in Figs.~1(b) and 1(c), respectively. The light pulse first propagates in the medium for $t < 400~\Gamma^{-1}$. It is stored in the medium for $400~\Gamma^{-1} < t < 800~\Gamma^{-1}$ and converted into a SLP at $t = 800~\Gamma^{-1}$. The SLP sustains for a period of $400~\Gamma^{-1}$ and is gradually broadened by a diffusion process. At $t = 1200~\Gamma^{-1}$ the SLP is converted back to a slowly propagating pulse.

\begin{figure}[t]
\includegraphics[width=8.25cm]{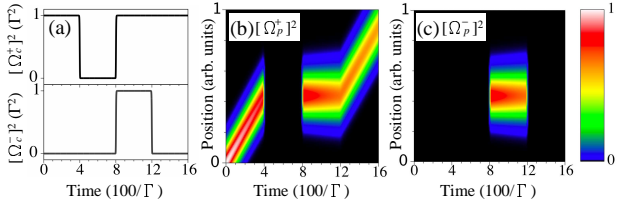}
\caption{(Color online) Numerical simulation of the formation of a SLP. The Gaussian shaped incident probe pulse has a $1/e$ full width of about $190/\Gamma$. (a) Timing diagram of the forward (black line) and backward (gray line) coupling fields. (b) and (c) $(\Omega^+_p)^2$ and $(\Omega^-_p)^2$ as functions of time (horizontal axis) and position (vertical axis), respectively. Color indicates the amplitude. The parameters are $\alpha = 900$ and $\gamma = 0$.}
\end{figure}

In the derivation of Eqs.~(\ref{two1})-(\ref{two5}), we do not require that $k^+_c = k^-_c$. This implies that the wavelength of the forward fields can be very different from that of the backward fields. To allow a more general treatment, we consider that the ground states $|1\rangle$ and $|2\rangle$ are coupled to different excited states $|3\rangle$ and $|4\rangle$ in the forward and backward directions, respectively. Equations~(\ref{two1})-(\ref{two5}) are still valid, except that $\rho^-_{31}$, $\Gamma$, and $\alpha$ in Eqs.~(\ref{two1}), (\ref{two3}), and (\ref{two5}) are replaced by $\rho^-_{41}$, $\Gamma'$, and $\alpha'$. $\Gamma'$ denotes the spontaneous decay rate of state $|4\rangle$ and $\alpha'$ denotes the optical density of the transition $|1\rangle\rightarrow|4\rangle$. The numerical calculation based on these more comprehensive equations also shows the formation of a SLP, as long as $\Omega^+_p/\Omega^+_c=\Omega^-_p/\Omega^-_c$. Even for $k^+_c$ very different from $k^-_c$, a SLP can still be created. This suggests that the standing wave formed at $k^+_c \approx k^-_c$, i.e., a Bragg grating is not necessary to produce a SLP. The formation of a SLP can however be explained for arbitrary $k^{\pm}_c$ by considering that the same ground state coherence $\rho_{21}$ is shared by the forward and backward fields in the balanced four-wave mixing (FWM) processes, $\omega_{p+} - \omega_{c+} + \omega_{c-} \rightarrow \omega_{p-}$ and $\omega_{p-} - \omega_{c-} + \omega_{c+} \rightarrow  \omega_{p+}$. Equations~(\ref{two2}) and (\ref{two3}) with the same $\rho_{21}$ can increase or decrease $\rho^+_{31}$ and $\rho^-_{41}$. As seen from Eqs.~(\ref{two4}) and (\ref{two5}), increment or decrement of $\rho^+_{31}$ and $\rho^-_{41}$ then affect $\Omega^+_p$ and $\Omega^-_p$, respectively. In the adiabatic regime in which $\partial \rho^+_{31} /\partial t$, $(\Gamma/2) \rho^+_{31}$, $\partial \rho^-_{41} /\partial t$,  and $(\Gamma'/2) \rho^-_{41}$ are negligible in Eqs.~(\ref{two2}) and (\ref{two3}), the solution of the equations is $\Omega^+_p /\Omega^+_c = -\rho_{21} = \Omega^-_p /\Omega^-_c$. In the case of $\Omega^+_c = \Omega^-_c$, $\Omega^+_p(z,t)$ must be the same as $\Omega^-_p(z,t)$ everywhere and always, i.e., a SLP is created. Our point of view of the SLP is in agreement with the theory of matched pulses \cite{HarrisMP} as well as with the theoretical finding that SLPs can be described in terms of unique dark state polaritons \cite{ZOU08}.

The experiments were performed in cold $^{87}$Rb atoms produced by a magneto-optical trap (MOT). There were $1.1 \times 10^9$ atoms in the MOT. The atom cloud had a cigar-like shape with a size of about $10 \times 2.5 \times 2.5$~mm$^3$. Details of our MOT setup and the production of the cigar-shaped atom cloud have been discussed elsewhere \cite{CigarMot}. Figure~2 shows the experimental setup and the transition diagrams. The probe and coupling fields were circularly polarized with right helicity. The two counter-propagating coupling beams had a $e^{-2}$ full width of 6.0~mm and interacted with all the trapped atoms. The probe beam was focused to a $e^{-2}$ full width of 0.27~mm and propagated along the major axis of the cigar-shaped atom cloud. The probe and coupling beams intersected at an angle of about $0.3^{\circ}$. Two avalanche photodiodes (APD, Hamamatsu C5460, photoelectric sensitivity $1.5 \times 10^6$~V/W, rise time 36 ns) simultaneously detected the probe transmissions in the forward and backward directions. After propagating through the atoms, the two coupling beams were blocked in order to reduce their influence on the signals detected by the two APDs. Parts of the forward- and backward-propagating probe beams were also blocked and the collection efficiencies of the two APDs were different. Other experimental details can be found in Ref.~\cite{OurXpm}.

\begin{figure}[t]
\includegraphics[width=8.25cm]{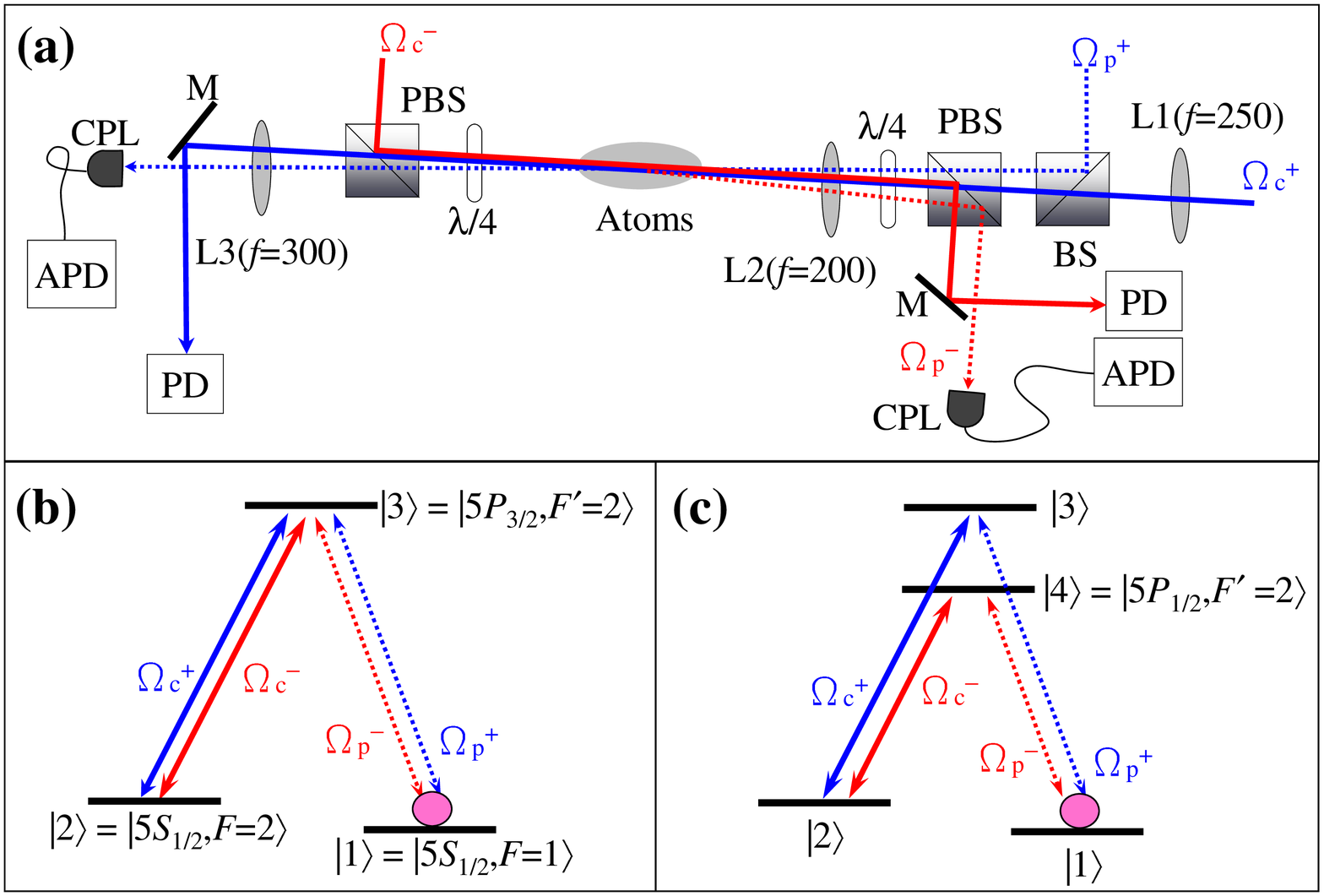}
\caption{(Color online) Experimental setup and coupling schemes. (a) Schematic experimental setup. BS, beam splitter cube; PBS, polarizing beam splitter cube; $\lambda$/4, quarter-wave plate; L1-L3, lenses ($f$ indicates the focal length in units of mm); CPL: optical fiber coupler; M, mirror; PD, photodetector; APD, avalanche photodetctor. We used L2 to focus the probe beam onto the atom cloud. L1 transformed the coupling beam into a plane wave in the region of the atom cloud. L3 collimates the transmitted probe beam. (b) Coupling scheme for monochromatic SLPs. (c) Coupling scheme for bichromatic SLPs.}
\end{figure}

First, we demonstrate the creation of a SLP in laser-cooled atoms, for coupling fields of the same wavelength around 780~nm (Fig.~3). The transition diagram is shown in Fig.~2(b). Several measurements were performed beforehand to adjust the experimental parameters. Figures~3(a) and 3(b) show the experimental data (solid lines) and theoretical predictions (dashed lines) of slow light and storage of light with the probe and coupling fields only applied in the forward direction, respectively. We adjusted the intensity of the coupling field such that nearly the entire Gaussian probe pulse was stored in the medium. By comparison of the experimental data and the theoretical prediction, we were able to determine the optical density, $\alpha$, forward coupling Rabi frequency, $\Omega^+_c$, and the relaxation rate of the ground-state coherence, $\gamma$, in the experiment. In Fig.~3(c) we show experimental and numerical data for the case of retrieving a previously forward propagating probe pulse with the backward coupling field after a storage period. The red dashed line is the theoretical prediction calculated with the parameters $\alpha$, $\Omega^+_c$, and $\gamma$ as determined from Figs.~3(a) and 3(b) and $\Omega^-_c\approx\Omega^+_c$. Because the collection efficiencies of the two APDs in the forward and backward directions were different, the theoretical data in Fig.~3(c) is scaled down by a factor of 0.42 in order to match the signal amplitude of the experimental data. Then, by comparison of the experimental and theoretical data we adjusted the backward coupling intensity such that $\Omega^+_c \approx \Omega^-_c$ as required for the creation of a SLP.

\begin{figure}[t]
\includegraphics[width=8.25cm]{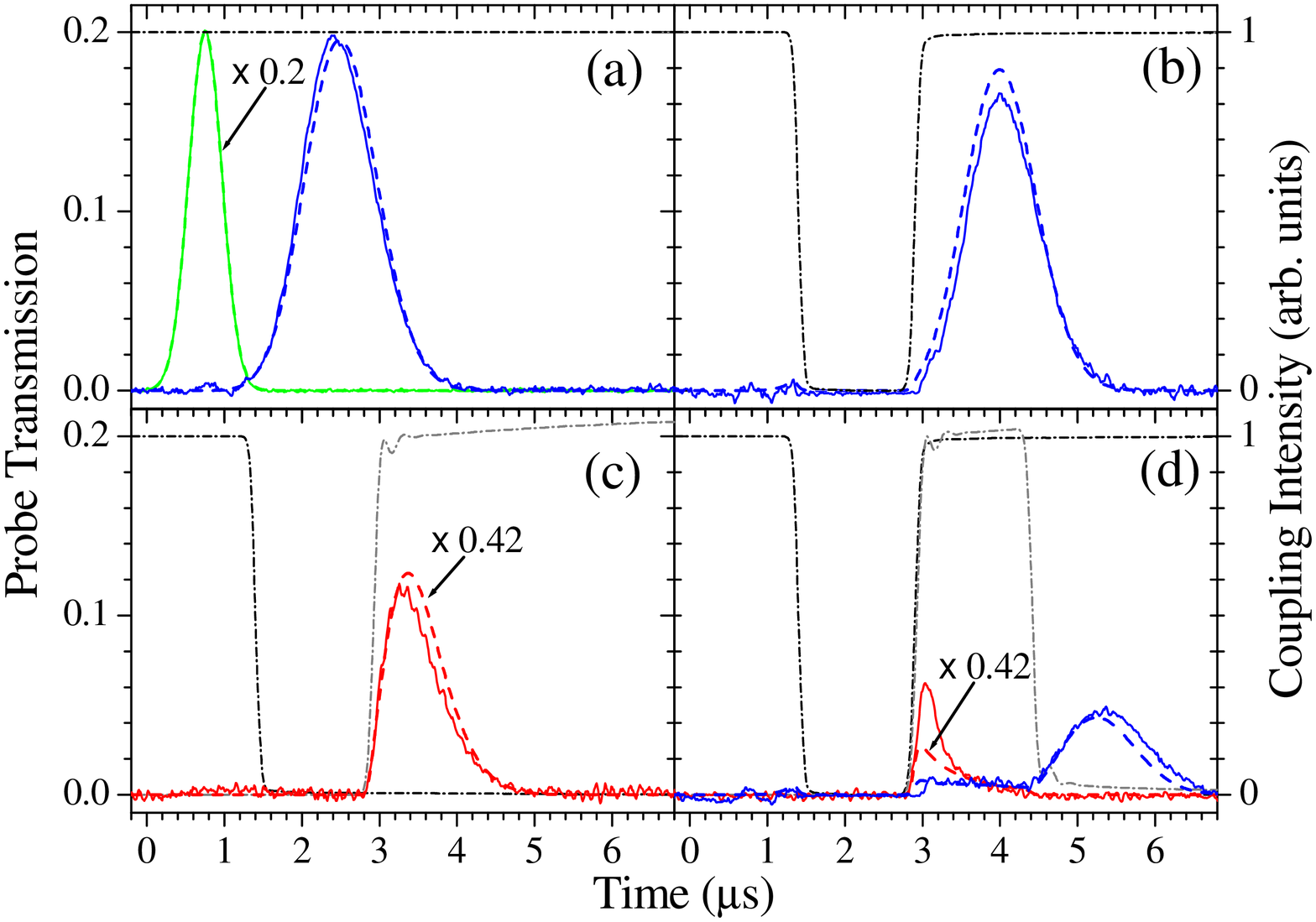}
\caption{(Color online) Creation of a monochromatic SLP. (a) Slow light and (b) storage of light in the forward direction only. (c) Storage of the forward-propagating probe pulse and retrieval in the backward direction. (d) Demonstration of the creation of a SLP in a gas of laser-cooled atoms. Solid lines show the experimental data. Dashed lines represent the theoretical predictions for the probe transmissions. Dashed-dotted lines show the measured temporal evolution of the forward (black) and backward (grey) coupling fields. We used hyperbolic tangent functions in the numerical simulations to model the evolution of the coupling fields that resembled the experimental data very well. The input probe pulse (green solid and dashed lines) is shown for reference in (a) and was scaled down by a factor of 0.2. Blue and red represent the forward and backward output probe signals. All probe signals are normalized with respect to the peak of the input probe pulse. The red dashed lines in (c) and (d) are rescaled by a factor of 0.42 to account for the different detection efficiencies in the forward and backward directions. All plots have the same horizontal and vertical scales. The parameters of the numerical simulation are $\Omega^+_c = \Omega^-_c = 0.77~\Gamma$ (peak amplitude), $\alpha = 39$ and $\gamma = 7.0 \times 10^{-4}~\Gamma$, where $\Gamma=2\pi\times 5.9$~MHz.}
\end{figure}

Figure~3(d) finally demonstrates that we created a SLP. The theoretical predictions shown in Fig.~3(d) are calculated with the $\alpha$, $\Omega^+_c$, $\Omega^-_c$, and $\gamma$ determined by Figs.~3(a), 3(b) and 3(c). Both forward and backward coupling fields were turned on simultaneously to form a standing wave during the retrieval. The two coupling frequencies differed slightly by 2.5 MHz. Because the optical density of the medium is not large enough, the probe signals leaked out of the medium in the forward and backward directions while the SLP was established. Such leakage has also been observed in the first experimental observation of a SLP \cite{SLP1}. By turning off the backward coupling field at $t \approx 4.5$ $\mu$s, the SLP was converted back to a slowly propagating pulse which left the medium in the forward direction. The qualitative agreement between the experimental data and the theoretical predictions is satisfactory. The pulse visible for $t>4.5~\mu$s represents the remaining energy of the initial probe pulse after a SLP duration of $1.5~\mu$s. We maximized this retrieved pulse by adjusting the time of turning off the forward coupling field, i.e., the spatial distribution of the ground-state coherence in the medium. Such adjustment affected the leakage in the forward and backward directions during the SLP period as the spatial distribution of the SLP is determined by that of the ground state coherence. A larger variation of the spatial distribution of a SLP at a boundary of the medium results in a larger leakage out of that boundary. For the theoretical data shown in Fig.~3(d) the peak of the probe pulse was stored 9~\% of the medium length away from the medium center closer to the input boundary. Currently we attribute the quantitative discrepancy between the experimental data and theoretical calculations to either a non-symmetrical density distribution of the atom cloud with respect to its center, or to reducing an actual two -or three-dimensional system to an one-dimensional system for the calculation. We plan to further study this issue in an upcoming publication.

In order to demonstrate that a Bragg grating, i.e., a standing wave coupling field is not necessary for the creation of a SLP, we changed the wavelength of the backward coupling field from 780~nm to 795~nm. This field drove the transition shown in Fig.~2(c). Because of the large difference between the wavelengths of 780~nm and 795~nm, the forward and backward coupling fields did not form a standing wave. Instead they formed a quasi-standing wave with a velocity of $c/105$ that moves the Bragg grating a distance of about 300 meters during the measurement time. We repeated the same measurements as those in Fig.~3 and employed the same method to determine $\alpha$, $\Omega^+_c$, $\Omega^-_c$, and $\gamma$. Figure~4(a) shows the storage of the forward-propagating probe pulse with the forward coupling field at 780~nm and the retrieval in the backward direction with the backward coupling field at 795~nm. Experimental and theoretical data are represented in the same way as in Fig.~3. The retrieved pulse has a larger amplitude and slightly different shape as compared to the data shown in Fig.~3(c). We believe this is due to a different spatial distribution of the ground state coherence during the retrieval as the relative timing of the coupling fields was slightly different as compared to Fig.~3(c).

\begin{figure}[t]
\includegraphics[width=8.25cm]{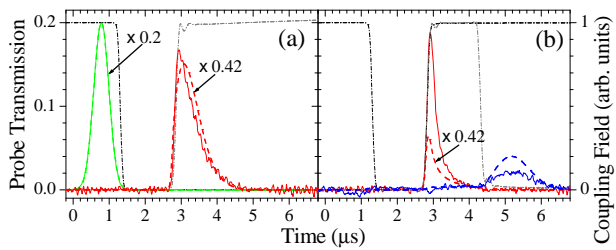}
\caption{(Color online) Creation of a bichromatic SLP. (a) Storage of a forward-propagating probe pulse with a forward coupling field at 780~nm and retrieval in the backward direction with a backward coupling field at 795~nm. (b) Demonstration of the creation of a bichromatic SLP without a Bragg grating. The legends and color codes are the same as those in Fig.~3. The input probe pulse has been scaled down by a factor of 0.2. Also the theoretical data has been scaled down by a factor of 0.42 as in Fig.~3(c) and 3(d). The parameters of the numerical simulation are $\Omega^+_c = 0.75~\Gamma$ and $\Omega^-_c = 0.77~\Gamma$ (peak amplitude), $\alpha = 39$ and $\gamma = 4.6 \times 10^{-4}~\Gamma$.}
\end{figure}

Figure~4(b) demonstrates the creation of a SLP without a Bragg grating. The typical leakage for SLPs in media with a rather low optical density is observed as for the measurement in Fig.~3(d). The measured signal at $t> 4.5$ $\mu$, which indicates the remaining probe energy after a SLP duration of $1.5~\mu$s, is smaller than the theoretical prediction. We attribute this to an increased relaxation rate of the ground-state coherence during the formation of the SLP, because the laser field at 795~nm was not phase-locked to the fields at 780~nm. Also, as the multi-order quarter-wave plates used in the experiment were designed for 780~nm, the 795~nm backward coupling field had a worse circular polarization than the coupling field in the forward direction. For the numerical results shown in Fig.~4 we neglected the difference between the spontaneous decay rates of the excited states $|5P_{3/2},F'=2\rangle$ and $|5P_{1/2},F'=2\rangle$ and the different absorption cross sections, as the differences are only about 5~\% and 4~\%, respectively. We believe that SLPs can not only be created with two, but with an unlimited number of different wavelengths. All that is required, is that the corresponding FWM processes are balanced and share the same ground state coherence.

\begin{figure}[t]
\includegraphics[width=8.25cm]{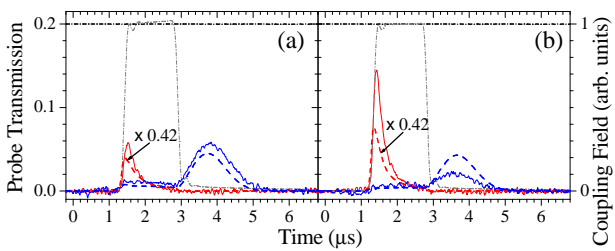}
\caption{(Color online) Creation of SLPs without storage. (a) Monochromatic SLP at 780~nm. (b) Bichromatic SLP at 780~nm and 795~nm. The legends, color codes, and calculation parameters are the same as those in Figs.~3 and 4.}
\end{figure}

Finally, we demonstrate experimentally that a SLP can be also created directly from a propagating probe pulse without prior storage. This is achieved by turning on the coupling field in the backward direction while the probe pulse is propagating through the medium in the forward direction. Using the backward coupling fields at 780~nm and 795~nm, respectively, Figs.~5(a) and 5(b) show the formation of SLPs without the process of light storage. These signals are very similar to those in Figs.~3(d) and 4(b). The results suggest that once the backward coupling field is switched on, it immediately utilizes the existing ground-state coherence of the DSP propagating through the medium in the forward direction to create a backward-propagating probe pulse, i.e., DSP. The amplitude of the backward-propagating probe pulse increases until the two probe pulses have the same amplitude everywhere, i.e. a SLP is created. Therefore, the process of light storage is not relevant for the creation of a SLP.

With these results we hope to have presented a more general view on the phenomenon of SLPs. As a result of balanced four-wave mixing processes that share the same ground state coherence, they fit well into the picture of multi-component DSPs \cite{AML06}. From this point of view SLPs seem to be so far the most general result from the concept of the DSPs introduced by Fleischhauer and Lukin \cite{FL00}. By choosing the number, relative direction and relative intensities of the coupling fields, light pulses can be efficiently manipulated - from multichromatic SLPs over slowly propagating light pulses to stored photonic information. The fact that these can be easily converted among each other by external fields with a high degree of freedom makes it a very promising tool for the manipulation of quantum information and nonlinear optical processes.

This work was supported by the National Science Council of Taiwan under Grants No. 95-2112-M-007-039-MY3 and No. 97-2628-M-007-018.

\begin{thebibliography}{99}
\bibitem{EitRv1}
S. E. Harris, Phys. Today {\bf 50}, 36 (1997).
\bibitem{EitRv2}
M. Fleischhauer, A. Imamo\v{g}lu, and J. P. Marangos, Rev. Mod. Phys. {\bf 77}, 633 (2005).
\bibitem{StopLtTh}
M. Fleischhauer and M. D. Lukin, Phys. Rev. Lett. {\bf 84}, 5094 (2000).
\bibitem{HauStopLt}
C. Liu, Z. Dutton, C. H. Behroozi, and L. V.  Hau, Nature (London) {\bf 409}, 490 (2001).
\bibitem{WalsworthStopLt}
D. F. Phillips, A. Fleischhauer, A. Mair, R. L. Walsworth, and M. D.  Lukin, Phys. Rev. Lett. {\bf 86}, 783 (2001).
\bibitem{SLP3}
A. Andr\'{e} and M. D. Lukin, Phys. Rev. Lett. {\bf 89}, 143602 (2002).
\bibitem{SLP1}
M. Bajcsy, A. S.  Zibrov, and M. D. Lukin, Nature (London) {\bf 426}, 638 (2003).
\bibitem{HarrisMP}
S. E. Harris, Phys. Rev. Lett. {\bf 70}, 552 (1993).
\bibitem{ZOU08}
F. E. Zimmer, J. Otterbach, R. G. Unanyan, B. W. Shore, and M. Fleischhauer, Phys. Rev. A {\bf 77}, 063823 (2008).
\bibitem{CigarMot}
Y. W. Lin, H. C. Chou, P. P. Dwivedi, Y. C. Chen, and I. A. Yu, Opt. Express {\bf 16}, 3753 (2008).
\bibitem{OurXpm}
Y. F. Chen, C. Y. Wang, S. H. Wang, and I. A. Yu, Phys. Rev. Lett. {\bf 96}, 043603 (2006).
\bibitem{AML06}
J. Appel, K.-P. Marzlin, and A. I. Lvovsky, Phys. Rev. A {\bf 73}, 013804 (2006).
\bibitem{FL00}
M. Fleischhauer and M. D. Lukin, Phys. Rev. Lett. {\bf 84}, 5094 (2000).
\end{thebibliography}
\end{document}